# NU:BRIEF – A Privacy-aware Newsletter Personalization Engine for Publishers

Ernesto Diaz-Aviles
recsyslabs
University College Dublin
Ireland
ernesto@recsyslabs.com

Claudia Orellana-Rodriguez
recsyslabs
University College Dublin
Ireland
claudia@recsyslabs.com

Igor Brigadir
recsyslabs
University College Dublin
Ireland
igor@recsyslabs.com

Reshma Narayanan Kutty
recsyslabs
University College Dublin
Ireland
reshma@recsyslabs.com

## ABSTRACT

Newsletters have (re-) emerged as a powerful tool for publishers to engage with their readers directly and more effectively. Despite the diversity in their audiences, publishers' newsletters remain largely a one-size-fits-all offering, which is suboptimal. In this paper, we present NU:BRIEF, a web application for publishers that enables them to personalize their newsletters without harvesting personal data. Personalized newsletters build a habit and become a great conversion tool for publishers, providing an alternative readers-generated revenue model to a declining ad/clickbait-centered business model.

**Demo:** https://demo.nubrief.com/md03PaAJSwXMegL5BbKpQlArK3elb3hDUglcHodx4gE=/

**Explainer video:** https://www.youtube.com/watch?v=AUZGuyPJYH4

## 1 INTRODUCTION

Newsletters are an attractive way for publishers to quickly build an audience and convert people into paying subscribers within weeks. For publishers, newsletters have become a viable revenue stream and an alternative to a declining click-bait business model.

For example, the New York Times has about 15 million subscribers across its 71 newsletters compared to 6.69 million digital-only subscriptions. In total, readers opened more than 3.6 billion newsletter emails from the publisher in 2020. The New York Times reports that digital revenue overtook print revenue, and digital subscriptions was its largest revenue stream. [6, 13]

However, the accelerated growth observed by major outlets such as the New York Times requires large amounts of resources for newsletter development, such as, an interdisciplinary team of technical staff, editorial specialists, and project managers. For small to medium publishers or independent writers with a limited budget, it is nearly impossible to tap on resources like these to maintain a diverse choice of newsletters. Their option is to settle on the better-than-nothing alternative to send the same newsletter to their diverse audience. This is suboptimal and clearly not on a growth path as the one reported by major media publishers.

Small to mid size publishers, however, do not have the legacy of a heavy machinery of old-line publishers, but have a better chance to innovate faster and start operating more like a digital product and technology company. One can observe the emergence of Neo-Media players serving and delighting audiences neglected by old-media that forgot their readers in exchange for minimal commission on click-bait ads served alongside their content. One key characteristic of these new players is that they are more respectful of reader privacy as a consequence of the relationship and trust they are looking to establish with their readers, who are increasingly privacy-conscious. In addition, publishers do not want the risk and liability of collecting and storing personal information given the intense scrutiny from data regulators.

NU:BRIEF, the tool we introduce in this work, enables publishers to monetize quality journalism by personalizing the newsletter experience for readers without the risk of collecting personal identifiable information, thereby moving away from a volatile advertising business. NU:BRIEF uses AI based on Machine Learning to automatically segment the publishers' audience into cohorts based on their interests and to produce a newsletter for each cohort with interesting articles tailored to the users' taste in each of these groups, without harvesting personal identifiable information or selling it to third parties.

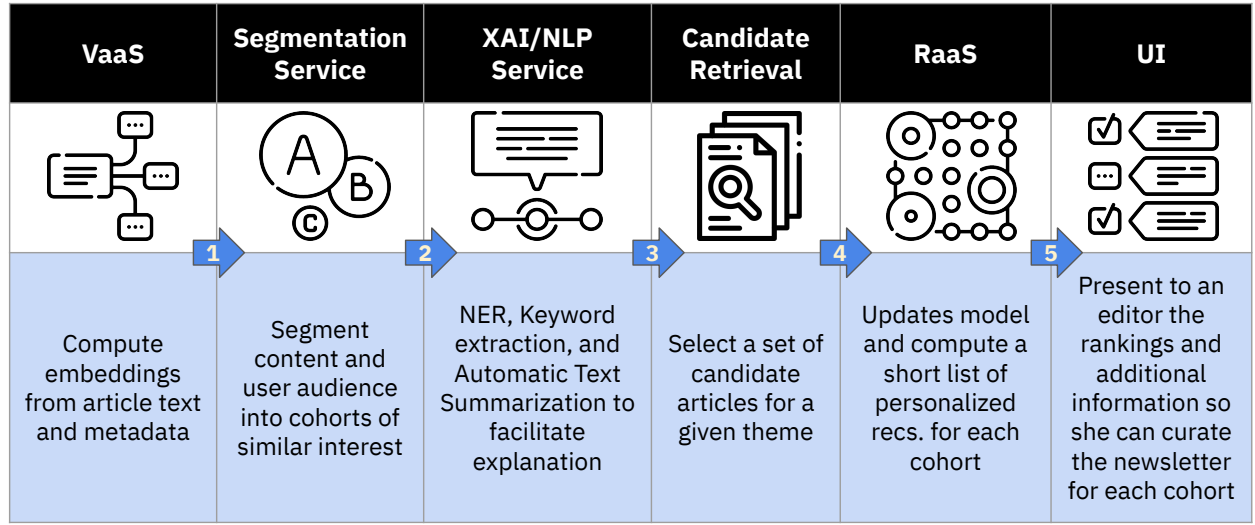

Figure 1: NU:BRIEF pipeline overview.[1]

In the rest of the paper, we introduce user stories to give context on how publishers are currently using NU:BRIEF to achieve their goals, we provide a technical overview of the system implementation, and present preliminary results on the online tests conducted.

## USER STORIES

Let us introduce two user stories to illustrate the scenarios in which NU:BRIEF can assist publishers to personalize their newsletters.

*User Story 1.* "As an editor, I want to be able to select a short list of interesting articles from a set of candidates about a theme of interest, so I write an introduction about the theme and include my article selection in our weekly newsletter."

*User Story 2.* "As an editor, I want to be able to rank (with the most relevant first) the articles we wrote this week to better inform and make our audience happy, so I can include them in our weekly newsletter."

## 2 NU:BRIEF OVERVIEW

Figure 1 shows NU:BRIEF's main pipeline. We detail each component as follows.

***Vector as a Service (VaaS).*** VaaS is responsible for computing vector representations, i.e., embeddings, from the text content of articles and its metadata. These embeddings are latent features extracted automatically using pre-trained NLP large language models based on transformers [3, 7, 12]. The standard flow is that articles from publishers using NU:BRIEF are initially processed offline during publisher onboarding. The articles are also indexed and the embedding is stored. New articles are processed periodically during the day, e.g., every 15m, to keep a cache system and index up to date.

***Segmentation Service.*** This component is responsible for discovering, initially from the content, a set of cohorts of interest. The rationale is that publishers write articles for a diverse audience and the topics of their production reflect the diversity of their target readers. We use k-means clustering [9] to this end and the embeddings output by the VaaS component as input. We have found that for publishers currently using NU:BRIEF a typical number of clusters (cohorts) $k$ ranges between 4 and 6, which is determined based on clustering metrics such as silhouette coefficient and a qualitative assessment.

***XAI/NLP Service.*** The goal of this component is to enrich the articles with additional information that can help editors, and readers, understand why a particular article is recommended. To this end, it extracts named entities, keywords, and automatic summaries from the articles. NU:BRIEF also computes embeddings using VaaS for these additional text elements and caches them to be used in later stages of the pipeline.

NU:BRIEF computes two automatic summaries, one *abstractive* and one *extractive* using transformers-based NLP models [8, 15]. In abstractive summarization the machine summarizes in its 'own' words what the article is about using language generation techniques, that is, the output concise text might not appear exactly in the input article. In contrast, extractive summarization identifies salient information in the article, which is then extracted and grouped together to form a concise summary. Both summaries are presented to the editor in the user interface.

***Candidate Retrieval Service.*** This service is responsible for indexing articles, their metadata, and embeddings. It is implemented as a search engine that assists, in production, the selection of a subset of candidate articles relevant to a specific theme or context.

---
[1]Icon illustrations designed by Freepik https://www.freepik.com.

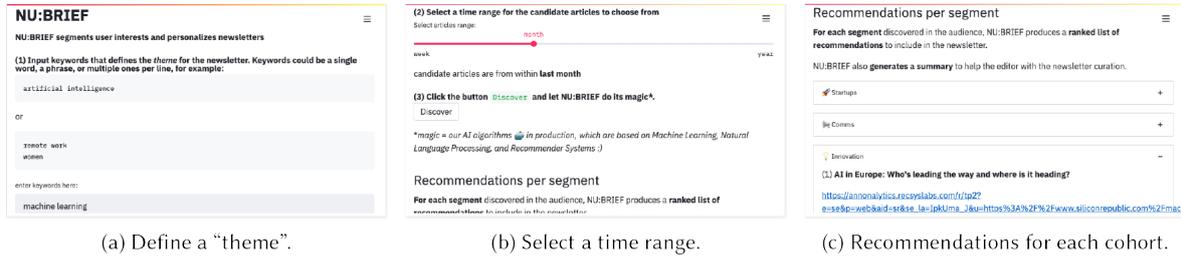

(a) Define a "theme".  (b) Select a time range.  (c) Recommendations for each cohort.

Figure 2: NU:BRIEF User Interface. Personalized newsletters in three basic steps.

These candidate articles are the ones to be re-ranked by the RaaS component to compute the personalized recommendations. This component is critical during inference since the cardinality of the candidates subset is significantly smaller than the whole set of articles indexed.

***Privacy-aware Recommender as a Service (RaaS).*** One of our design principles for NU:BRIEF is to develop a recommender system service that is private-by-default, compliant with privacy regulations, and offers a high quality personalization. As part of this goal, NU:BRIEF computes an aggregate representation of a user's taste (*taste vector*) based on their history interactions (e.g., clicks) with the recommended content. Differential privacy [5], e.g., data perturbation, is applied in this step for additional privacy guarantees.

The core recommender system engine uses a hybrid approach of content-based [1], matrix factorization collaborative filtering [4, 10] and k-nearest neighbors (k-NN) [14]. In our experience, deep learning architectures have served us well in production for learning representations from text and in the XAI pipeline, e.g., automatic summarization, but we have found that relatively simple recommender system algorithms, e.g., embedding based models and dot product operations on those embeddings to compute scores, have proven very practical to deploy in production and have performed well in live tests. In addition, offline evaluations we have conducted show that these methods have not been inferior to neural-based models for recommendation, which is in line with recent studies, e.g., [2, 11]. We are currently experimenting with Deep and Cross Network [17] architectures and with attention mechanisms [16] to improve model interpretability guided by high-level journalistic features.

As a side product of our recommender technology, we generate privacy-preserving analytics which can inform how cohorts of users interact with content in different contexts and scenarios, without singling out or exposing any individual.

***User Interface.*** NU:BRIEF user interface is shown in Figure 2. It is designed so that the newsletter editor fulfills the user stories presented in Section 1. There are three basic steps: (i) define newsletter theme using phrases or keywords, (ii) specify a time range for candidate articles to consider, and (iii) receive personalized recommendations per cohort of interest. NU:BRIEF versions in production also offer an export to html functionality for publishers so they can include the selected articles recommended into their newsletter distribution system.

## 3 EVALUATION

Over the first half of 2021, we have conducted a pilot test in a private beta with small-to-medium publishers whose content covers topics ranging from technology, STEM, startups, local news, and food recipes. These publishers serve a combined audience of more than 43K subscribers for their newsletters. In total, NU:BRIEF pipeline has processed 135K documents published by the pilot publishers.

The main lessons learned from A/B tests conducted using NU:BRIEF are as follows. (i) Personalized rankings per cohort computed by NU:BRIEF perform at the same level as the ones curated by a human editor. That is, newsletters produced based on NU:BRIEF rankings achieve opening rates and CTR that are equivalents to the ones achieved by the ones curated entirely manually by a journalist. (ii) Publishers report that time to curate a newsletter drops from one hour to 10 minutes when using NU:BRIEF (80% time savings). (iii) The conversion rate from newsletter readers to paying subscribers increases with the personalization options offered by NU:BRIEF.

## 4 CONCLUSION

In this work, we have given an overview of our approach, NU:BRIEF, towards building a private-by-default, personalized newsletter recommender engine. NU:BRIEF produces a ranked list of recommendations per segment (cohort) of common interest to assist editors in newsletter curation. As private-by-default recommender systems do not exist off-the-shelf, our work will allow publishers to provide personalized newsletters without violating privacy regulations, and allow users to find content they enjoy without sacrificing their private data.

## ACKNOWLEDGMENTS

This work is supported by Enterprise Ireland grant no. CS20191123, a project co-funded by the European Regional Development Fund (ERDF) under Ireland's European Structural and Investment Funds Programme 2014-2020.